\documentclass[a4paper,12pt]{article}
\usepackage[a4paper,text={16.8cm,22.4cm}]{geometry}
\usepackage{amsmath,amssymb,bm,graphicx}
\allowdisplaybreaks 
\def\rd{\mathrm{d}}
\def\nb{\bar{n}}
\def\cdt{ \hspace{-0.1em} \cdot \hspace{-0.1em} }

\begin{document}

\begin{titlepage}

\begin{flushright}
August 11, 2010
\end{flushright}

\vspace{0.7cm}
\begin{center}
\Large\bf
\boldmath
The gluon jet function at two-loop order
\unboldmath
\end{center}

\vspace{0.2cm}
\begin{center}
{\sc Thomas Becher and Guido Bell}\\
\vspace{0.4cm}
{\sl Institute for Theoretical Physics \\
University of Bern\\
Sidlerstrasse 5, 3012 Bern, Switzerland}
\end{center}

\vspace{1.0cm}
\begin{abstract}
\vspace{0.2cm}
\noindent 
The inclusive gluon jet function is evaluated at two-loop accuracy. This function is relevant for resummations of large perturbative logarithms in collider processes involving low-mass gluon jets. The jet function corresponds to the imaginary part of the gluon propagator in light-cone gauge, which is adopted for the calculation. In addition to the leading jet function, the power-suppressed two-gluon jet functions are given and their renormalization is discussed.
\end{abstract}
\vfil

\end{titlepage}

\section{The gluon jet function}

Cross sections for the production of low-mass jets at large momentum transfer often factorize into hard, jet and soft functions \cite{Sen:1981sd,Kidonakis:1998bk}. While the hard functions encode the physics associated with the large momentum transfer $Q$, the jet functions describe the propagation of collinear partons inside the jets. The soft functions mediate low energy interactions between jets and have a low characteristic scale $M^2/Q$, where $M$ is the typical invariant mass of the jets. 

Soft-Collinear Effective Theory (SCET) implements this factorization on the operator level \cite{Bauer:2000yr,Bauer:2001yt,Beneke:2002ph}. In this framework the hard functions appear as Wilson coefficients of operators built out of soft and collinear fields. At leading power, the operators do not contain soft fields and the soft interactions can be decoupled from the effective Lagrangian. After the decoupling, the soft functions are given by matrix elements of light-like Wilson lines along the directions of large energy flow. The matrix elements of the collinear fields associated with the final state define the jet functions. The collinear fields associated with the initial state at hadron colliders define (generalized) parton distribution functions, also called beam functions \cite{Stewart:2009yx}. By solving the renormalization group (RG) equations of the  hard, jet and soft functions, one can resum contributions to jet cross sections which are logarithmically enhanced by ratios of the different scales.

Most of the early applications of SCET were concerned with processes mediated by electro\-weak currents with only two directions of large energy flow. Such processes involve only quark jets.  The corresponding jet function was computed to two-loop order already some time ago \cite{Becher:2006qw}. The gluon jet function first appeared in computations of radiative $\Upsilon$ decays \cite{Fleming:2002rv,Fleming:2002sr,Fleming:2003gt}, but more generally, it appears in any process involving gluon jets. This includes any jet process at hadron colliders and the production of three or more jets at $e^+e^-$ machines. The jet function is given by the vacuum matrix element of two gluon fields, 
\begin{equation} \label{jetfef0}
\int \rd^4 x\,e^{i p x}\, \langle  0 |\, {\cal A}^a_\mu( x) {\cal A}^b_\nu(0)\,   | 0 \rangle \\
 =  \sum_X (2\pi)^4\, \delta^{(4)}(p - p_X)\, \langle  0 |\, {\cal A}^a_\mu(0) \,   | X \rangle \langle  X |\,  {\cal A}^b_\nu(0)   | 0 \rangle \,.
\end{equation}
Loosely speaking, it measures the probability that a gluon field produces a jet of particles with momentum $p$ from the vacuum. The field ${\cal A}_\mu(x)$ is the gluon field in light-cone gauge. It is related to the field $A_\mu(x)$ in a general gauge via
\begin{equation}
\label{eq:collfields}
{\cal A}^\mu(x)= {{\cal A}^a}^\mu( x) t_a = W^\dagger(x)\,[iD^\mu W(x)] \, .
\end{equation}
The light-like ($n^2=0$) Wilson lines
\begin{equation}\label{eq:Whc}
W(x) = {\rm\bf P}\exp\left(ig_s\int_{-\infty}^0\!ds\, n\cdt A(x+s n) \right)
\end{equation}
ensure that the jet function is invariant under gauge transformations which vanish for $x\to \infty$.  The symbol ${\bf{P}}$ indicates path ordering, and the conjugate Wilson line $W^\dagger$ is defined with the opposite ordering prescription. In SCET the jet function is defined as a matrix element of collinear fields which describe the propagation of energetic particles with small invariant masses $p^2 \ll (n\cdot p)^2$. However, after decoupling the soft fields, the collinear Lagrangian is equivalent to the QCD Lagrangian. For the computation of the matrix element (\ref{jetfef0}) we may therefore use QCD Feynman rules.

The one-loop result for the gluon jet function was given in \cite{Becher:2009th}, where it was used to perform soft-gluon resummation for direct photon production at next-to-next-to-leading logarithmic (NNLL) accuracy. Without too much effort it should be feasible to push the logarithmic accuracy for vector boson production processes even one order higher. Based on the results of \cite{Becher:2009cu,Becher:2009qa} (see also \cite{Gardi:2009qi,Dixon:2009ur}), the necessary three-loop anomalous dimensions were derived in \cite{Becher:2009th}. The two-loop hard functions are also known and can be extracted from the results of \cite{Garland:2001tf,Garland:2002ak}. The only missing ingredients to perform a N$^3$LL resummation for $W$, $Z$ and photon production at large transverse momentum are the two-loop results for the soft function and the gluon jet function. In the present paper, we compute the gluon jet function at two-loop order.

In addition to the inclusive jet function considered in this work, it is also interesting to study jet functions that incorporate kinematical restrictions on the final state $X$ which appears in (\ref{jetfef0}). Such jet functions become relevant when a jet-algorithm is applied to the final state. The one-loop quark jet function for Sterman-Weinberg jets has recently been computed in \cite{Jouttenus:2009ns}, and both quark and gluon jet functions for general recombination and cone algorithms have been worked out in \cite{Ellis:2010rw} (for related work cf.~\cite{Trott:2006bk,Cheung:2009sg}).

The most general parameterization of the gluon jet function consistent with the identity $n \cdot {\cal A}(x)=0$ and invariance under a rescaling of the light-cone vector $n_\mu$ reads
\begin{multline} \label{jetfef1}
\int \rd^4 x\,e^{i p x}\, \langle  0 |\, {\cal A}^a_\mu( x) {\cal A}^b_\nu(0)\,   | 0 \rangle \\
 =  \delta^{ab} \, g_s^2\, \theta(p^0)\, \left[\left(-g_{\mu\nu}+\frac{n_\mu p_\nu+ p_\mu n_\nu}{n\cdot p}  \right) J_g(p^2) + \frac{n_\mu n_\nu}{(n\cdot p)^2} K_g(p^2)  \right] \, ,
\end{multline}
where the strong coupling constant $g_s$ on the right-hand side is the bare coupling.
 In the construction of SCET one introduces a conjugate light-cone vector $\bar{n}_\mu$ with $\bar{n}\cdot n=2$ and decomposes the gluon field as
\begin{equation}\label{eq:split}
{\cal A}^\mu(x)= {\cal A}_\perp^\mu(x) + \frac{n^\mu}{2} \bar{n} \cdot {\cal A}(x)\, ,
\end{equation}
with $\bar{n}\cdot {\cal A}_\perp(x)= n\cdot {\cal A}_\perp(x) =0$. The jet function $J_g(p^2)$ then corresponds to the matrix element of two transverse fields ${\cal A}_\perp^\mu(x)$. Since the component $ \bar{n} \cdot {\cal A}(x)$ is suppressed in the limit $p^2 \ll (n\cdot p)^2$, only the function $J_g(p^2)$ is relevant in leading-power factorization theorems. However, independence of the reference vector $\bar{n}_\mu$ ensures that the field components always appear in the combination (\ref{eq:split}) and for completeness we will compute the full two-point function, including the power-suppressed piece  $K_g(p^2)$.

In the following section, we will discuss the technical aspects of the two-loop computation and present the bare, unrenormalized, two-loop results. As a check of our calculation, we have performed the computation both in Feynman and light-cone gauge. The renormalization of the jet functions is discussed in Section \ref{sec:renorm}. As an interesting aside, we discuss the choice of the operator basis necessary to renormalize the power-suppressed jet function $K_g(p^2)$.

\section{Two-loop calculation}

In order to calculate the jet function, it is convenient to rewrite (\ref{jetfef1}) as the imaginary part of the time-ordered product 
\begin{equation}\label{calJ}
 \int \rd^4x\, e^{i p x}  \langle 0 | \,  {\bm T}\left\{ {\cal A}^a_\mu(x) {\cal A}^b_\nu(0) \right \} \, | 0
\rangle  = \delta^{ab} \, g_s^2 \left[\left(-g_{\mu\nu}+\frac{n_\mu p_\nu+ p_\mu n_\nu}{n\cdot p}  \right) {\cal J}_g(p^2) + \frac{n_\mu n_\nu}{(n\cdot p)^2} {\cal K}_g(p^2)  \right]
\end{equation}
with
\begin{align}
{J}_g(p^2) &= \frac{1}{\pi} {\rm Im} \left[ i {\cal J}_g(p^2) \right],
&
{K}_g(p^2) &= \frac{1}{\pi} {\rm Im} \left[ i {\cal K}_g(p^2) \right] \,.
\end{align}
This form shows that the jet function is given by the imaginary part of the gluon propagator in light-cone gauge $n \cdot A(x) = 0$, where the Wilson lines become trivial and ${\cal A}^\mu(x) = g_s A^\mu(x)$. For dimensional reasons, the function ${\cal J}_g(p^2)\sim (p^2)^{-1}$, while ${\cal K}_g(p^2)\sim (p^2)^0$. Thus only ${\cal J}_g(p^2)$ receives contributions from single-particle intermediate states, while ${\cal K}_g(p^2)$ starts at one-loop and has a cut for $p^2>0$ but no pole.

Since the Wilson lines are absent in light-cone gauge, it is convenient to use this gauge in the computation of the jet function. The free gluon propagator in light-cone gauge is given by
\begin{equation}\label{eq:gluonprop}
\frac{i}{p^2+i0} \left[-g_{\mu\nu}+\frac{n_\mu p_\nu+n_\nu p_\mu}{n\cdot p} \right] \,.
\end{equation}
Note that we are not adopting the Mandelstam-Leibbrandt (ML) prescription to regulate the $n\cdot p \to 0$ singularity. The ML prescription \cite{Mandelstam:1982cb,Leibbrandt:1983pj}

\begin{equation}
\frac{1}{n\cdot p} \to  \frac{\bar{n}\cdot p }{n\cdot p\,\bar{n}\cdot p+i 0}
\end{equation} 
cures the collinear singularity in the propagator, but in SCET this singularity has a physical meaning. The Wilson line $W(x)$ and the associated light-cone propagators arise from expanding QCD diagrams around the large-energy limit and the choice of the $i0$-prescription is dictated by the QCD diagrams. The loop integrals contributing to the jet function are unambiguously defined in dimensional regularization. They depend on a single four momentum $p$. If they involve $k$ light-cone propagators they scale as $(n\cdot p)^{-k}$. Since the dependence on $n\cdot p$ is analytic, the final result of the integral is independent of the sign of the $i 0$-prescription adopted for the light-cone propagator.

\begin{figure}[t!]
\begin{center}
\includegraphics[width=0.6\textwidth]{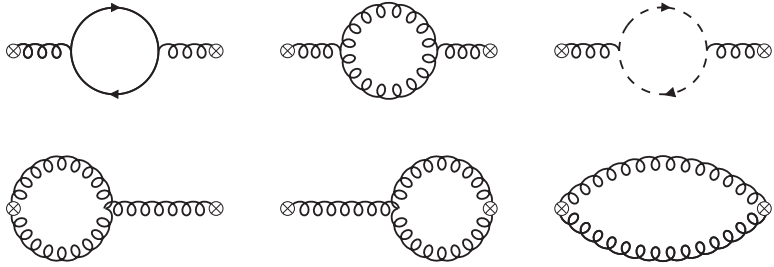}
\end{center}
\caption{Diagrams contributing to the gluon jet function at one-loop. In light-cone gauge the ghost contribution, represented by the third diagram, as well as the diagrams in the second row, which involve emissions from the Wilson lines (denoted by the crosses), are absent. \label{fig:oneLoopGraphs}}
\end{figure}

\begin{figure}[t!]
\begin{center}
\begin{tabular}{c}
\includegraphics[width=0.9\textwidth]{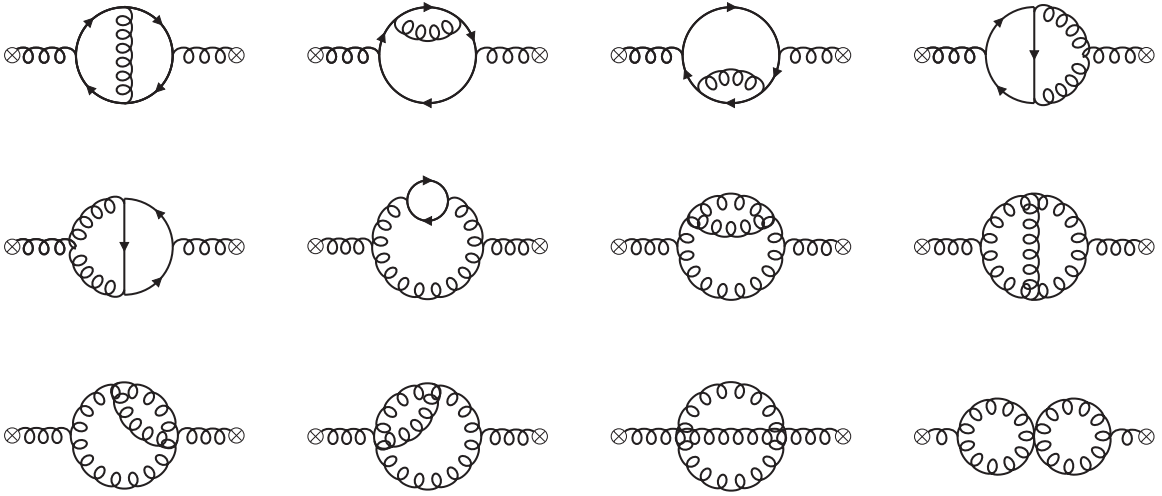}
\end{tabular}
\end{center}
\caption{
Two-loop diagrams that arise in the light-cone gauge calculation. 
\label{fig:twoLoopGraphs}}
\end{figure}

As a check of our result we performed the calculation both in light-cone and Feynman gauge. Both gauges offer some advantages, but overall the light-cone gauge computation is more efficient. The number of diagrams in this gauge is significantly smaller, because no Wilson line diagrams and no ghost-loop contributions need to be included. At one-loop the light-cone gauge calculation reduces for instance to the first two diagrams in Figure~\ref{fig:oneLoopGraphs}. At two-loop order, the light-cone gauge calculation involves the diagrams shown in Figure~\ref{fig:twoLoopGraphs}, while the number of diagrams is again more than doubled in Feynman gauge.
Furthermore, in light-cone gauge, standard tools for QCD computations can be used, e.g.\  we generate the diagrams with FeynArts \cite{Hahn:2000kx}. The only disadvantage of light-cone gauge is the proliferation of light-cone denominators. In the diagrams in Figure \ref{fig:twoLoopGraphs} each gluon propagator has an associated light-cone denominator, while in Feynman gauge such denominators only arise from Wilson lines. 

The most general two-loop integral that appears in the light-cone gauge calculation has the form
\begin{multline}
\label{eq:twoloopintGeneral}
   \int\!d^dk\,\int\!d^dl\, 
   \frac{1}{\left(k^2\right)^{a_1} \left(l^2\right)^{a_2}  \left[(k-l)^2\right]^{a_3}
 \left[(k+p)^2\right]^{b_1}
         \left[(l+p)^2\right]^{b_2} \left[(k+l+p)^2\right]^{b_3}} \\
        \times \frac{1}{\left(\nb\cdot k\right)^{A_1} \left(\nb\cdot l\right)^{A_2}
 \left(\nb\cdot k-\nb\cdot l\right)^{A_3}  \left(\nb\cdot k+\nb\cdot p \right)^{B_1} \left(\nb\cdot l+\nb\cdot p\right)^{B_2} \left(\nb\cdot k+\nb\cdot l+\nb\cdot p\right)^{B_3}  }     
\end{multline}
This class of integrals contains more master integrals than those which appeared in the two-loop computation of the quark jet function \cite{Becher:2006qw}. Nevertheless, after using partial-fractioning identities and symmetry relations, no additional master integrals remain after solving the integration-by-parts relations \cite{Chetyrkin:1981qh} for the integrals which actually appear in the diagrams. To solve the integration-by-parts relations we use Laporta's \cite{Laporta:2001dd} algorithm as implemented in the public codes AIR \cite{Anastasiou:2004vj} and FIRE \cite{Smirnov:2008iw} as well as an independent code written by one of us.

After combining the irreducible two-loop diagrams with one-particle reducible ones, we find that both calculations yield the same bare result. We obtain
   \begin{multline}
   i{\cal J}_g(p^2) =  \frac{1}{-p^2-i0} \left\{
   1 + \frac{Z_\alpha \alpha_s}{(4\pi)} \left(\frac{\mu^2}{-p^2-i0}\right)^\epsilon J_1(\epsilon) 
  \right. \\ \left.
   +  \frac{Z_\alpha^2 \alpha_s^2}{(4\pi)^2} \left(\frac{\mu^2}{-p^2-i0}\right)^{2\epsilon} \left[   C_A^2
 J_{AA}(\epsilon) +  C_A n_f T_F J_{Af}(\epsilon) +  C_F n_f T_F J_{Ff} +  n_f^2 T_F^2 J_{ff} (\epsilon)  \right]
 \right\} \,,
   \end{multline}
where $\alpha_s$ refers to the $\overline{\rm MS}$ coupling constant, which is related to the bare coupling constant $\alpha_s^0$ via  $Z_\alpha\, \alpha_s \,\mu^{2\epsilon} = e^{-\epsilon \gamma_E}(4\pi)^\epsilon \alpha_s^0$ with $Z_\alpha  = 1-\beta_0 \alpha_s/(4\pi \epsilon)$ and $\beta_0 = 11/3 \,C_A - 4/3\, T_F n_f$. The one-loop coefficient is \cite{Becher:2009th}
   \begin{equation}
 J_1(\epsilon) =   \left[C_A \left(\frac{3}{\epsilon}-\frac{9}{4} \right)-n_f T_F \right]
 e^{\epsilon \gamma_E} \frac{8 \Gamma(2-\epsilon)^2\,\Gamma(\epsilon)}{\Gamma(4-2\epsilon)}\, ,  
\end{equation}
and the two-loop coefficients are found to be
 \begin{align}
 J_{AA}(\epsilon) &= \frac{8}{\epsilon^4}+\frac{55}{3 \epsilon ^3} +\frac{1}{\epsilon ^2}\left(-\frac{\pi^2}{3}+
 \frac{152}{3}\right)+\frac{1}{\epsilon}\left(-\frac{184 \zeta_3}{3}-\frac{11 \pi ^2}{6} +\frac{3638}{27} \right) 
 \nonumber\\
 & \;\;\;\;\;-\frac{23 \pi ^4}{180}-\frac{1496 \zeta_3}{9}-\frac{161 \pi^2}{27}+\frac{57415}{162} \,,\nonumber\\
    J_{Af}(\epsilon) &= -\frac{20}{3 \epsilon ^3}-\frac{188}{9 \epsilon^2}+\frac{1}{\epsilon }\left(\frac{2 \pi
 ^2}{3}-\frac{536}{9} \right)+\frac{400 \zeta_3}{9}+\frac{74 \pi^2}{27}-\frac{12880}{81} \,,\\
   J_{Ff}(\epsilon) &= -\frac{2}{\epsilon }+ 16 \zeta_3-\frac{55}{3}  \, ,\nonumber \\
    J_{ff}(\epsilon) &= \frac{16}{9 \epsilon ^2}+\frac{160}{27 \epsilon }-\frac{8 \pi^2}{27}+16  \, .\nonumber
 \end{align}
The result for the bare function ${\cal K}_g$ takes the form
  \begin{equation}
   i{\cal K}_g(p^2) = \frac{Z_\alpha \alpha_s}{(4\pi)} \left(\frac{\mu^2}{-p^2-i0}\right)^\epsilon K_1(\epsilon) 
   +  \frac{Z_\alpha^2 \alpha_s^2}{(4\pi)^2} \left(\frac{\mu^2}{-p^2-i0}\right)^{2\epsilon} \left[   C_A^2
 K_{AA}(\epsilon) +  C_A n_f T_F K_{Af}(\epsilon) \right]  \,. 
   \end{equation}
Only a single color-structure appears at one-loop
   \begin{equation}
 K_1(\epsilon) =   -C_A\, e^{ \epsilon \gamma_E  } \frac{4 \Gamma (2-\epsilon ) \Gamma
   (-\epsilon ) \Gamma (\epsilon )}{\Gamma (2-2 \epsilon )} \, ,
   \end{equation}
and only two structures at two-loop level
 \begin{align}
 K_{AA}(\epsilon) &= \frac{8}{\epsilon^4}+\frac{19}{\epsilon ^3}+\frac{148}{3 \epsilon ^2}+\frac{1}{\epsilon
 }\left(-\frac{151 \zeta_3}{3}-\frac{7 \pi^2}{6}+\frac{1082}{9} \right)+\frac{17 \pi^4}{72}-\frac{428
 \zeta_3}{3}-\frac{41 \pi ^2}{9}+\frac{7672}{27} \,,\nonumber\\
    K_{Af}(\epsilon) &=-\frac{4}{\epsilon ^3}-\frac{38}{3 \epsilon^2}-\frac{298}{9 \epsilon }+ \frac{44
 \zeta_3}{3}+\frac{13 \pi ^2}{9}-\frac{2198}{27}  \, . 
    \end{align}
The bare jet functions $J^{\rm bare}_g(p^2)$ and $K^{\rm bare}_g(p^2)$ then follow by taking the imaginary part using
 \begin{equation}
 \frac{1}{\pi}{\rm Im}\left[ (-p^2-i 0 )^a \right] = -\theta(p^2) \frac{\sin(\pi a)}{\pi} (p^2)^a\,.
 \end{equation}
 The function $J^{\rm bare}_g(p^2)$ is a distribution in $p^2$ whose explicit form is obtained after expanding
 \begin{equation} \label{eq:star}
\frac{1}{p^2} \left(\frac{p^2}{\mu^2}\right)^{-\epsilon} = -\frac{1}{\epsilon} \delta(p^2) + \sum_{n=0}^{\infty} \frac{(-\epsilon)^n}{n!} \left[ \frac{\ln^n(\frac{p^2}{\mu^2})}{p^2} \right]^{[\mu^2]}_* \,.
 \end{equation}
The star-distributions are generalizations of plus-distributions to dimensionful variables, their definition can be found in \cite{Bosch:2004th}.

\section{Renormalization\label{sec:renorm}}

\subsection{The leading jet function $J_g(p^2)$}

In momentum space the leading jet function renormalizes via the convolution
\begin{equation}
 J_g(p^2,\mu) = \int_0^{p^2} \,dp'^2 Z_{J_g} (p^2-p'^2,\mu)\, J^{\rm bare}_g(p'^2) \,.
 \end{equation}
Both the jet function and the $Z$-factor are distribution valued, see (\ref{eq:star}). For this reason it is more convenient to perform the renormalization in Laplace space, where it is multiplicative. The Laplace transformed function $\widetilde{j}_g$ is defined as
\begin{equation}
\widetilde{j}_g\left(\ln \frac{Q^2}{\mu^2},\mu\right) = \int_0^\infty d p^2 e^{-\nu p^2} J_g(p^2,\mu)\, ,
\quad\text{ with }  
\quad
\nu= \frac{1}{Q^2 e^{\gamma_E}}\, .
\end{equation}
The function $\widetilde{j}_g$ is also what is needed to perform soft-gluon resummation in the momentum space formalism of \cite{Becher:2006nr}. It fulfils the RG equation 
\begin{align} \label{eq:RG}
\frac{\rd}{\rd \ln\mu}  \, \widetilde{j}_g\left(\ln \frac{Q^2}{\mu^2},\mu\right)
&= \left[ - 2 \Gamma^A_{\mathrm{cusp}} \ln  \frac{Q^2}{\mu^2}   -2 \gamma^{J_g} \right] \widetilde{j}_g \left(\ln \frac{Q^2}{\mu^2},\mu\right) \,,
\end{align}
where $\Gamma^A_{\mathrm{cusp}}$ denotes the cusp anomalous dimension, which is known at the three-loop level \cite{Moch:2004pa}. The jet anomalous dimension $\gamma^{J_g}$ was inferred to three loops in \cite{Becher:2009th} using RG invariance of the direct photon-production cross section and the three-loop results for the hard anomalous dimensions \cite{Becher:2009cu,Becher:2009qa}, the quark jet anomalous dimension \cite{Becher:2006mr}, and the Casimir scaling property of the soft function. Expanding  the anomalous dimensions as $\Gamma^A_{\mathrm{cusp}}= \sum_{n=0}^\infty \Gamma^A_n (\frac{\alpha_s}{4\pi})^{n+1}$ and $\gamma^{J_g} = \sum_{n=0}^\infty \gamma^{J_g}_n \,(\frac{\alpha_s}{4\pi})^{n+1}$, the two-loop solution to the RG equation (\ref{eq:RG}) takes the form
\begin{align} \label{eq:RGsol}
\widetilde{j}_g (L,\mu ) &= 1 + \left( \frac{\alpha_s}{4 \pi} \right)
\left[ \Gamma^A_0 \frac{L^2}{2} + \gamma_0^{J_g} L + c_1^{J_g} \right] + \left(
\frac{\alpha_s}{4 \pi} \right)^2 \left[ \left( \Gamma_0^A \right)^2
\frac{L^4}{8} + \left( - \beta_0 + 3 \gamma_0^{J_g} \right) \Gamma_0^A
\frac{L^3}{6} \right. \nonumber\\
& \left. + \left( \Gamma_1^{J_g} + (\gamma_0^{J_g})^2 - \beta_0 \gamma_0^{J_g} + c_1^{J_g}  \Gamma_0^A \right) \frac{L^2}{2} 
+ (\gamma_1^{J_g} + \gamma_0^{J_g} c_1^{J_g} - \beta_0  c_1^{J_g}) L + c_2^{J_g} \right]\,.
\end{align}
The coefficients entering at this order are 
\begin{align}
\Gamma_0^A &= 4 C_A \,, &
\Gamma_1^A &= 4 C_A \left[ C_A \left( \frac{67}{9} - \frac{\pi^2}{3} \right) -
\frac{20}{9} T_F n_f \right]\,, \\
\gamma_0^{J_g} &= - \beta_0\,, & \gamma_1^{J_g} &= C_A^2 \left( - \frac{1096}{27} + \frac{11 \pi^2}{9} + 16  \zeta_3 \right) + C_A n_f T_F \left( \frac{368}{27} - \frac{4 \pi^2}{9}  \right) + 4 C_F T_F n_f \,.\nonumber 
\end{align}
Together with the one-loop coefficient \cite{Becher:2009th}
\begin{align}
c^{J_g}_1 &= C_A \left( \frac{67}{9} - \frac{2\pi^2}{3} \right) -  \frac{20}{9} T_F n_f  \,,
\end{align}
the solution (\ref{eq:RGsol}) completely determines the two-loop jet function up to the constant $c_2^{J_g}$, which we compute in this work.

In Laplace space the jet function renormalizes multiplicatively, $\widetilde{j}_g = Z_{\tilde j_g} \widetilde{j}_g^{\rm bare}$, and $Z_{\tilde j_g}$ fulfils the same RG equation (\ref{eq:RG}) as the renormalized jet function. Solving this equation \cite{Becher:2009cu,Becher:2009qa}, one derives the following expression for the logarithm of the $Z$-factor:
\begin{multline} 
\ln{ Z}_{\tilde j_g} = \frac{\alpha _s}{4 \pi } \left[-\frac{ \Gamma^{\rm A}_0 }{\epsilon ^2}
+\frac{1}{
   \epsilon }\left(\Gamma^{\rm A}_0
    \ln\frac{Q^2}{\mu^2}+\gamma^{J_g}_0\right)\right]
 \\+ \left(\frac{\alpha _s}{4 \pi }\right)^2
  \Bigg[  \frac{3 \beta _0 \Gamma^{\rm A}_0 }{4 \epsilon ^3} 
 -\frac{\beta _0}{2 \epsilon ^2}\left(\Gamma^{\rm A}_0  \ln \frac{Q^2}{\mu^2} + \gamma^{J_g}_0 \right) 
  -\frac{\Gamma^{\rm A}_1 }{4\epsilon ^2} 
+\frac{1}{2 \epsilon}\left( \Gamma^{\rm A}_1  \ln\frac{Q^2}{\mu^2} + \gamma^{J_g}_1 \right)
 \Bigg] \,. 
   \end{multline}
Since all the necessary anomalous dimensions are known, the $Z$-factor is completely determined at the two-loop level. The cancellation of all divergences $1/\epsilon^n$, for $n=1\dots 4$, in the renormalized result provides a strong check of our calculation. We finally obtain for the non-logarithmic two-loop coefficient 
 \begin{equation}  
   \begin{aligned}
 c^{J_g}_2 = &C_A^2 \left(\frac{20215}{162}-\frac{362 \pi ^2}{27}-\frac{88 \zeta_3}{3}+\frac{17 \pi
   ^4}{36}\right) \\
   & +C_A n_f T_F \left(-\frac{1520}{27}+\frac{134 \pi
   ^2}{27}-\frac{16 \zeta_3}{3}\right) \\
   &+   C_F n_f T_F\left(-\frac{55}{3}+16 \zeta_3\right)
 +  n_f^2 T_F^2\left(\frac{400}{81}-\frac{8 \pi ^2}{27}\right) \,.
   \end{aligned}
   \end{equation}
  Numerically, for $n_f=5$ flavors, this yields
  \begin{multline}
\widetilde{j}_g (L,\mu ) = 1 + \left( \frac{\alpha_s}{4 \pi} \right) \left(-2.961 - 7.667\, L + 6 L^2\, \right) \\
+\left( \frac{\alpha_s}{4 \pi} \right)^2 \left( -58.58 + 44.39\, L + 82.46\, L^2 - 61.33\, L^3 + 18\, L^4 \right) \,,
\end{multline}
which may be compared to the quark case 
 \begin{multline}
\widetilde{j}_q (L,\mu ) = 1 + \left( \frac{\alpha_s}{4 \pi} \right) \left( 0.560 - 4 L + 2.667 L^2\right) \\
+\left( \frac{\alpha_s}{4 \pi} \right)^2 \left(-36.34 + 32.14\, L + 43.25\, L^2 - 17.48\,L^3 + 3.56\,L^4 \right)\,.
\end{multline}
Not surprisingly, the corrections are larger in the gluon case. On the other hand, for $\alpha_s =0.1$ they are numerically small. For $|L|<1$, the two-loop corrections amount to $ \pm 0.4\%$. To gauge their phenomenological relevance, we have included the corrections into the resummed result for the direct photon cross section \cite{Becher:2009th}. At the Tevatron, and for $p_T=50\,{\rm  GeV}$, the change in the cross section due to the inclusion of the two-loop gluon jet function is around $+1\%$, while the two-loop corrections from the hard and soft functions (which are, however, only partially known) are both around $+5\%$.

\subsection{The subleading jet function $K_g(p^2)$}

While there exists only one physical operator at leading power, we need to include additional power-suppressed operators to perform the renormalization of the subleading jet function. This is obvious, since $K_g(p^2)$ vanishes at tree level, but is divergent at one-loop order. To discuss renormalization, it is convenient to work with the scalar operators
\begin{equation}
\begin{aligned}
O_{\cal J}(x) &= \frac{-g_{\mu\nu}}{d-2}  {\bm T}\!\left\{ {\cal A}^a_\mu(x)\, {\cal A}^b_\nu(0) \right \}\, ,\\
O_{\cal K}(x) &= {\bm T}\!\left\{ \partial\cdot {\cal A}^a(x)\, \partial\cdot{\cal A}^b(0) \right \} + \Box \,O_{\cal J}(x)  \, ,
\end{aligned}
\end{equation}
whose vacuum matrix elements give rise to the jet functions $J_g(p^2)$ and $K_g(p^2)$, respectively. In $d=4$ dimensions, $O_{\cal J}(x) \sim 1/x^2$ and $O_{\cal K}(x) \sim 1/x^4$, so the second operator is power-suppressed with respect to the first one. 

For our purposes it is sufficient to concentrate on power-suppressed two-gluon operators, which mix into operators with more than two fields, but not vice versa. Starting from the leading operator $O_{\cal J}$, infinitely many power-suppressed two-gluon operators can be constructed. Examples include
\begin{equation}
\begin{aligned}
O_{1}(x) &= -\Box \,O_{\cal J}(x) \, , &
O_{2}(x) &= \frac{4}{-x^2+i 0} \,O_{\cal J}(x) \,, &
O_{3}(x) &= 4\,\frac{\partial}{\partial x^2} \,O_{\cal J}(x) 
\,.
\end{aligned}
\end{equation}
In Laplace space one finds that the vacuum matrix elements of these operators are
\begin{equation}
\begin{aligned} \label{eq:ops:laplace}
{\widetilde j}_1(Q^2) & = e^{\gamma_E}\,Q^4  \frac{\partial }{\partial Q^2} \, {\widetilde j}_g(Q^2) =  {\cal O}(\alpha_s) \, ,\\  
{\widetilde j}_2(Q^2) & =  e^{\gamma_E}\, \big(Q^2\big)^\epsilon \int_0^{Q^2}\!\!\! {\rm d} {Q^\prime}^2 \big({Q^\prime}^2\big)^{-\epsilon}\, {\widetilde j}_g({Q^\prime}^2) =  \frac{e^{\gamma_E}\,Q^2}{1-\epsilon} + {\cal O}(\alpha_s)\, , \\
{\widetilde j}_3(Q^2)  & = e^{\gamma_E}\,Q^2\,{\widetilde j}_g(Q^2) = e^{\gamma_E}\,Q^2 + {\cal O}(\alpha_s) \, ,
\end{aligned}
\end{equation}
where ${\widetilde j}_g$ is the leading (bare) jet function. Notice that, at tree-level, ${\widetilde j}_1$ vanishes and the difference ${\widetilde j}_2-{\widetilde j}_3$ also vanishes in $d=4$ dimensions. Since the operators $O_{1}$, $O_{2}$ and $O_{3}$ have the same field content, including spin structure, their physical matrix elements cannot be distinguished and it is therefore sufficient to include an arbitrary linear combination with non-vanishing tree-level matrix element in the operator basis. If several of such operators are included, the renormalization scheme can always be chosen such that only one combination has a non-vanishing renormalized matrix element. In this sense the additional operators are similar to evanescent operators which appear in matching computations onto four-quark operators \cite{Buras:1989xd}. However, in contrast to this case, we are not forced to include the additional operators to obtain a basis which closes under renormalization.   

From (\ref{eq:ops:laplace}) it is obvious, that the simplest choice corresponds to only including the operator $O_{3}$. In momentum space, the corresponding jet function is just the integral over the leading jet function,
\begin{equation}
J_3(p^2) = \int_0^{p^2} {\rm d} {p^\prime}^2\, J_g({p^\prime}^2) \,.
\end{equation}
For completeness, let us also give the momentum space representation of $O_2$, which reads
\begin{equation}
J_2(p^2)= \frac{2}{d-2} \left\{ J_3(p^2)  +  \int_{p^2}^\infty \!\! {\rm d} {p^\prime}^2
 \left(\frac{{p^\prime}^2}{p^2}\right)^\frac{d-2}{2} \, J_g({p^\prime}^2) \, \right \} \, .
\end{equation}
Note that the integral in the second term is not ultraviolet convergent in $d=4$ dimensions and can thus not be performed on the level of the renormalized jet function $J_g(p^2,\mu)$. 
A similar behavior has been observed in convolution integrals relevant for power corrections to the process $\bar B \to X_s \gamma$ \cite{Benzke:2010js}. Subleading quark jet functions have recently been studied in \cite{Paz:2009ut}, where it was noticed that an operator basis which includes only the equivalent of $O_{1}$ is insufficient to perform renormalization. It was further speculated whether the operator basis closes under renormalization if $O_3$ is added. Our discussion makes it clear that this is the case, as far as additional operators related to the leading jet function are concerned. 

Let us now discuss the renormalization of the two subleading jet functions $\widetilde{k}_g$ and ${\widetilde j}_3$. The renormalized jet functions are obtained as
\begin{equation}
\left( \begin{matrix} {\widetilde j}_3 \\ \widetilde{k}_g \end{matrix} \right) = \left( \begin{matrix}  Z_{\tilde{j}_g} & 0 \\   Z_{kj} & Z_{kk}   \end{matrix} \right) \, \left( \begin{matrix} {\widetilde j}^{\rm bare}_3 \\ \widetilde{k}^{\rm bare}_g\end{matrix} \right) \,.
\end{equation}
The first row follows from the fact that ${\widetilde j}_3 = e^{\gamma_E} Q^2 {\widetilde j}_g$ and that the leading operator ${\widetilde j}_g$ cannot mix into the subleading operator. Of the two other coefficients only $Z_{kj} $ enters at one-loop order, and we find 
\begin{equation}
Z_{kj} = - \frac{C_A \alpha_s}{\pi}  \frac{1}{\epsilon} \, ,
\end{equation}
and
\begin{equation}
\widetilde k_g (Q^2,\mu) =  \frac{C_A \alpha_s}{\pi} Q^2 e^{\gamma_E} \left( 1- \ln\frac{Q^2}{\mu^2}\right) .
\end{equation}
To obtain the renormalized result at two loops, we would need to compute the renormalization factor $Z_{kk}$ at one-loop order. To separate out corrections associated with $Z_{kk}$  from the ones proportional to $Z_{kj}$, one would need to compute a matrix element which is sensitive to the spin of the gluon fields in the operators. The simplest possibility is to compute a one-gluon matrix element of the operator $O_{\cal K}$ at one-loop. 

\section{Conclusions}

We computed the gluon jet function at two-loop order. Performing the calculation in both Feynman and light-cone gauge, we argued that the latter leads to considerable simplifications. The current calculation yields one of the two missing ingredients to perform the N$^3$LL soft-gluon resummation for $W$, $Z$ and photon production at large transverse momentum. In addition, we computed the subleading two-gluon jet functions and discussed how to construct a minimal basis of power-suppressed operators which closes under renormalization.

\vspace{0.6cm}
{\em Acknowledgments:\/}
This work is supported in part by funds provided by the Schweizerischer Nationalfonds (SNF). The Albert Einstein Center for Fundamental Physics at the University of Bern is supported by the Innovations- und Kooperationsprojekt C-13 of the Schweizerische Universita\"atskonferenz (SUK/CRUS).


\newpage
 \begin{thebibliography}{99}  

\bibitem{Sen:1981sd}
 A.~Sen,
 Phys.\ Rev.\  D {\bf 24}, 3281 (1981).

\bibitem{Kidonakis:1998bk}
 N.~Kidonakis, G.~Oderda and G.~F.~Sterman,
 Nucl.\ Phys.\  B {\bf 525}, 299 (1998)
 [arXiv:hep-ph/9801268].

\bibitem{Bauer:2000yr}
 C.~W.~Bauer, S.~Fleming, D.~Pirjol and I.~W.~Stewart,
 Phys.\ Rev.\  D {\bf 63}, 114020 (2001)
 [arXiv:hep-ph/0011336].

\bibitem{Bauer:2001yt}
 C.~W.~Bauer, D.~Pirjol and I.~W.~Stewart,
 Phys.\ Rev.\  D {\bf 65}, 054022 (2002)
 [arXiv:hep-ph/0109045].

\bibitem{Beneke:2002ph}
 M.~Beneke, A.~P.~Chapovsky, M.~Diehl and T.~Feldmann,
 Nucl.\ Phys.\  B {\bf 643}, 431 (2002)
 [arXiv:hep-ph/0206152].

\bibitem{Stewart:2009yx}
 I.~W.~Stewart, F.~J.~Tackmann and W.~J.~Waalewijn,
 Phys.\ Rev.\  D {\bf 81}, 094035 (2010)
 [arXiv:0910.0467 [hep-ph]].

\bibitem{Becher:2006qw}
 T.~Becher and M.~Neubert,
 Phys.\ Lett.\  B {\bf 637}, 251 (2006)
 [arXiv:hep-ph/0603140].

\bibitem{Fleming:2002rv}
 S.~Fleming and A.~K.~Leibovich,
 Phys.\ Rev.\ Lett.\  {\bf 90}, 032001 (2003)
 [arXiv:hep-ph/0211303].

\bibitem{Fleming:2002sr}
 S.~Fleming and A.~K.~Leibovich,
 Phys.\ Rev.\  D {\bf 67}, 074035 (2003)
 [arXiv:hep-ph/0212094].

\bibitem{Fleming:2003gt}
 S.~Fleming, A.~K.~Leibovich and T.~Mehen,
 Phys.\ Rev.\  D {\bf 68}, 094011 (2003)
 [arXiv:hep-ph/0306139].

\bibitem{Becher:2009th}
 T.~Becher and M.~D.~Schwartz,
 JHEP {\bf 1002}, 040 (2010)
 [arXiv:0911.0681 [hep-ph]].

\bibitem{Becher:2009cu}
 T.~Becher and M.~Neubert,
 Phys.\ Rev.\ Lett.\  {\bf 102}, 162001 (2009)
 [arXiv:0901.0722 [hep-ph]].

\bibitem{Becher:2009qa}
 T.~Becher and M.~Neubert,
 JHEP {\bf 0906}, 081 (2009)
 [arXiv:0903.1126 [hep-ph]].
 
 
\bibitem{Gardi:2009qi}
  E.~Gardi and L.~Magnea,
  JHEP {\bf 0903}, 079 (2009)
  [arXiv:0901.1091 [hep-ph]].
  
\bibitem{Dixon:2009ur}
  L.~J.~Dixon, E.~Gardi and L.~Magnea,
  JHEP {\bf 1002}, 081 (2010)
  [arXiv:0910.3653 [hep-ph]].

\bibitem{Garland:2001tf}
 L.~W.~Garland, T.~Gehrmann, E.~W.~N.~Glover, A.~Koukoutsakis and E.~Remiddi,
 Nucl.\ Phys.\  B {\bf 627}, 107 (2002)
 [arXiv:hep-ph/0112081].

\bibitem{Garland:2002ak}
 L.~W.~Garland, T.~Gehrmann, E.~W.~N.~Glover, A.~Koukoutsakis and E.~Remiddi,
 Nucl.\ Phys.\  B {\bf 642}, 227 (2002)
 [arXiv:hep-ph/0206067].

\bibitem{Jouttenus:2009ns}
 T.~T.~Jouttenus,
 Phys.\ Rev.\  D {\bf 81}, 094017 (2010)
 [arXiv:0912.5509 [hep-ph]].

\bibitem{Ellis:2010rw}
 S.~D.~Ellis, C.~K.~Vermilion, J.~R.~Walsh, A.~Hornig and C.~Lee,
 arXiv:1001.0014 [hep-ph].

\bibitem{Trott:2006bk}
  M.~Trott,
  Phys.\ Rev.\  D {\bf 75}, 054011 (2007)
  [arXiv:hep-ph/0608300].

\bibitem{Cheung:2009sg}
  W.~Y.~Cheung, M.~Luke and S.~Zuberi,
  Phys.\ Rev.\  D {\bf 80}, 114021 (2009)
  [arXiv:0910.2479 [hep-ph]].

\bibitem{Mandelstam:1982cb}
 S.~Mandelstam,
 Nucl.\ Phys.\  B {\bf 213}, 149 (1983).

\bibitem{Leibbrandt:1983pj}
 G.~Leibbrandt,
 Phys.\ Rev.\  D {\bf 29}, 1699 (1984).

\bibitem{Hahn:2000kx}
 T.~Hahn,
 Comput.\ Phys.\ Commun.\  {\bf 140}, 418 (2001)
 [arXiv:hep-ph/0012260].

\bibitem{Chetyrkin:1981qh}
 K.~G.~Chetyrkin and F.~V.~Tkachov,
 Nucl.\ Phys.\  B {\bf 192}, 159 (1981).

\bibitem{Laporta:2001dd}
 S.~Laporta,
 Int.\ J.\ Mod.\ Phys.\  A {\bf 15}, 5087 (2000)
 [arXiv:hep-ph/0102033].

\bibitem{Anastasiou:2004vj}
 C.~Anastasiou and A.~Lazopoulos,
 JHEP {\bf 0407}, 046 (2004)
 [arXiv:hep-ph/0404258].

\bibitem{Smirnov:2008iw}
 A.~V.~Smirnov,
 JHEP {\bf 0810}, 107 (2008)
 [arXiv:0807.3243 [hep-ph]].

\bibitem{Bosch:2004th}
 S.~W.~Bosch, B.~O.~Lange, M.~Neubert and G.~Paz,
 Nucl.\ Phys.\  B {\bf 699}, 335 (2004)
 [arXiv:hep-ph/0402094].

\bibitem{Becher:2006nr}
 T.~Becher and M.~Neubert,
 Phys.\ Rev.\ Lett.\  {\bf 97}, 082001 (2006)
 [arXiv:hep-ph/0605050].

\bibitem{Moch:2004pa}
 S.~Moch, J.~A.~M.~Vermaseren and A.~Vogt,
 Nucl.\ Phys.\  B {\bf 688}, 101 (2004)
 [arXiv:hep-ph/0403192].

\bibitem{Becher:2006mr}
 T.~Becher, M.~Neubert and B.~D.~Pecjak,
 JHEP {\bf 0701}, 076 (2007)
 [arXiv:hep-ph/0607228].

\bibitem{Buras:1989xd}
  A.~J.~Buras and P.~H.~Weisz,
  Nucl.\ Phys.\  B {\bf 333}, 66 (1990).
  
  \bibitem{Benzke:2010js}
  M.~Benzke, S.~J.~Lee, M.~Neubert and G.~Paz,
  arXiv:1003.5012 [hep-ph].

 \bibitem{Paz:2009ut}
  G.~Paz,
  JHEP {\bf 0906}, 083 (2009)
  [arXiv:0903.3377 [hep-ph]].


  \end{thebibliography}
\end{document}